\documentclass[referee,sn-mathphys-num]{sn-jnl}

\usepackage{graphicx}%
\usepackage{multirow}%
\usepackage{amsmath,amssymb,amsfonts}%
\usepackage{amsthm}%
\usepackage{mathrsfs}%
\usepackage[title]{appendix}%
\usepackage{xcolor}%
\usepackage{textcomp}%
\usepackage{manyfoot}%
\usepackage{booktabs}%
\usepackage{algorithm}%
\usepackage{algorithmicx}%
\usepackage{algpseudocode}%
\usepackage{listings}%
\usepackage[textsize=tiny]{todonotes}

\usepackage{lineno}
\usepackage{graphicx} 

\begin{document}

\title[Article Title]{Diverse, Distinct, and Densely Packed DNA Droplets}
\maketitle
\centerline{\large Aria S. Chaderjian$^1$, Sam Wilken$^{1,2,3}$, Omar A. Saleh$^{1,2}$*}

\vspace{1cm}
\noindent{1. Physics Department, University of California, Santa Barbara, California, 93106, USA.}

\noindent{2. Materials Department, University of California, Santa Barbara, California, 93106, USA.}

\noindent{3. Current affiliation: Department Chemie, Johannes Gutenberg University - Mainz, Mainz, 55122, Germany.}



\noindent{*Corresponding author. E-mail: saleh@ucsb.edu }

\vspace{0.8cm}






\centerline{\textbf{Abstract}}
\noindent{The liquid-liquid phase separation of biomolecules is an important process for intracellular organization. Biomolecular sequence combinatorics leads to a large variety of proteins and nucleic acids which can interact to form a diversity of dense liquid (`condensate') phases. The relationship between sequence design and the diversity of the resultant phases is therefore of interest. Here, we explore this question using the DNA nanostar system which permits the creation of multi-phase condensate droplets through sequence engineering of the sticky end bonds that drive particle-particle attraction. We explore the theoretical limits of nanostar phase diversity, then experimentally demonstrate the ability to create 9 distinct, non-adhering nanostar phases that do not share components. We further study how thermal processing affects the morphology and dynamics of such a highly diverse condensate system. We particularly show that a rapid temperature quench leads to the formation of a densely packed 2-D layer of droplets that is transiently stabilized by caging effects enabled by the phase diversity, leading to glassy dynamics, such as slow coarsening and dynamic heterogeneity. Generally, our work provides experimental insight into the thermodynamics of phase separation of complex mixtures and demonstrates the rational engineering of complex, long-range, multi-phase droplet structures.}

\vspace{0.8cm}
\section{Introduction}
The interior of the living cell consists of a large diversity of biological macromolecules that are typically found in a demixed state. 
Recent advances have clarified that the cellular cytoplasm and nucleoplasm spontaneously compartmentalize into multiple membraneless sub-volumes that are dense in specific biomolecules while others are excluded~\cite{banani2017biomolecular}.
These compartments, often liquid-like, are generally termed condensates and form due to weak, multivalent interactions between biomolecules.
While non-equilibrium processes can be relevant to condensate formation, thermodynamic models have been successful in predicting condensate formation in \emph{in vitro} experiments, permitting the view of condensates as liquid-liquid phase-separated (LLPS) droplets \cite{jacobs2023theory}.

The complexity of the cell's biomolecular mixture raises the question of the maximal number of distinct, mutually immiscible condensates that can be formed from a set of biomolecules.
Recent theoretical work has investigated this, including both general investigations of the thermodynamics of multi-component mixtures \cite{mao2019phase,shrinivas2021phase,jacobs2021self,chen2023programmable,teixeira2024liquid} and more chemically specific studies that account for the nature of biomolecular interactions \cite{lin2017charge,chen2023multiphase,sanders2020competing,chew2023theromdynamic,harmon2018differential}.
Certain experimental works have shown that mixtures of two or three biomolecular components, if appropriately chosen, can form a similar number of phases \cite{Kaur2021sequence, kelley2021amphiphilic,arora2025chaperone,rai2023heterotypic,lu2025controlling,gupta2023bacterial,rana2024asymmetric,ye2024micropolarity,zhou2025multiphasic,fisher2020tunable,Feric2016coexisting}.

We argue that a DNA system is optimal for experimental realization of many diverse macromolecular phases due to the programmability and specificity of duplex hybridization.
Particularly, a system of DNA particles offers the ability to engineer specific pairwise interactions through variations in base sequence, allowing microscopic control of mixing thermodynamics.
The binding affinity of two short complementary DNA sequences is sensitive to single-base changes, allowing different sequences to have orthogonal binding modalities.
This creates a potentially large combinatorial spectrum of distinct interactions and, accordingly, a large potential diversity in the number of phases.
In contrast, the disordered proteins (IDPs) common to biological condensates have relatively limited interaction modalities, as encoded through residue-residue interactions \cite{li1997nature,borcherds2021how}; consequently, multi-phase condensates formed from IDPs tend to adhere and share components \cite{Kaur2021sequence, kelley2021amphiphilic,arora2025chaperone,rai2023heterotypic,lu2025controlling,gupta2023bacterial,rana2024asymmetric,ye2024micropolarity,zhou2025multiphasic,fisher2020tunable,Feric2016coexisting}. 

\begin{figure}
  \centering
  \includegraphics[width=8.8cm]{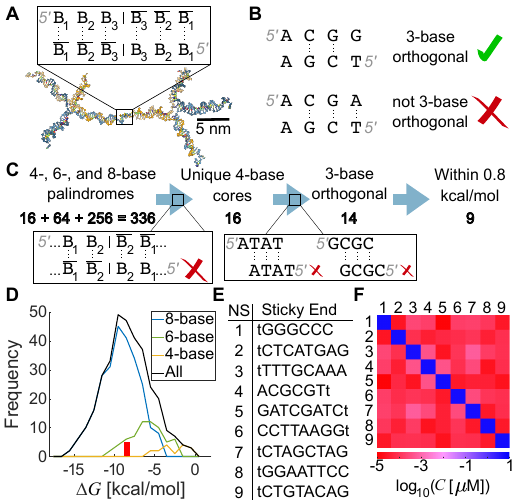}
  \caption{
  A) Schematic of two bound NSs, created using OxDNA \cite{snodin2015introducing}. A generic 6-base self-binding (palindromic) sticky end sequence is shown, with $\overline{B_i}$ representing the Watson-Crick complement of base $B_i$. The vertical line indicates the midway point of the palindrome. B) Examples of sequences that are, and are not, orthogonal at the level of 3 successive bases. C) Overview of the argument predicting the maximal size of a set of mutually orthogonal palindromic sticky ends that all have a similar binding energy; see text for details. D) Histogram of binding energies, $\Delta G$, of all 4-, 6-, and 8-base palindromes, calculated using the SantaLucia data set \cite{santalucia2004thermodynamics}.  E) A set of 9 sticky end sequences that satisfy the constraints indicated in (C), and that are explored experimentally here. The $\Delta G$ for these sequences (with unpaired T omitted) are shown as red lines in (D). Lowercase letters indicate the base intended to be unpaired upon NS-NS binding, to impart flexibility to the structure. F) An effective pairwise interaction matrix of the 9 sticky ends shown in (E), computed as the concentrations of all possible dimers in a multi-component mixture containing 12 $\mu$M of each of the 9 sticky ends shown in (E) with 500 mM NaCl and at 25$^\circ$C. 12 $\mu$M of a sticky end corresponds to 3 $\mu$M of a 4-armed NS, matching later experimental conditions. Computations performed using NUPACK \cite{zadeh2011nupack}.
}
  \label{fig:theory}
\end{figure}

These considerations raise the possibility, potentially unique to nucleic acid systems, of creating sets of many condensates that are completely physically distinct, with no significant sharing of components and no inter-phase adhesion, without the use of surfactants or other stabilizing mechanisms.
To explore this, we demonstrate the creation of diverse and distinct biomolecular liquids using the DNA nanostar system~\cite{biffi2013phase}.
Nanostars (NSs) are roughly 10 nm, branched, self-assembled DNA particles consisting of several double-stranded arms that emanate from a common junction (Fig.~\ref{fig:theory}A).
Each arm terminates in a single-stranded segment (a sticky end) whose sequence controls interactions between NSs through the Watson-Crick rules of DNA duplex formation.
NSs can be designed to exhibit homotypic interactions by choosing sticky end sequences to be self-binding, i.e. palindromic. 
Solutions of homotypic NSs are driven by sticky end binding to demix through an (LLPS) process with an upper critical solution temperature $T_c$ \cite{biffi2013phase} .
In appropriate conditions, homotypic NSs at $T<T_c$ will form micron-scale DNA droplets that sediment \cite{jeon2018salt}.
Further, prior works have shown that the use of appropriately designed sticky ends can lead multi-component NS mixtures to form two or three distinct phases \cite{Jeon2020sequence, sato2020sequence, Gong2022computational}.

We use the NS system to investigate the limits of condensate diversity, particularly testing the ability to create a large set of homotypic DNA phases that are mutually immiscible, do not share components, and are completely non-adhering.
We develop a theoretical argument, based on sequence combinatorics and DNA duplex thermodynamics, that predicts the maximum diversity of homotypic NS condensates given the constraints of palindrome structure, sequence orthogonality, and materials properties of the resulting phases.
Then, we carry out proof-of-principle experiments that realize 9 distinct, fully demixed, and completely non-adhering DNA droplet phases.
We finally demonstrate that appropriate processing of such a diverse NS mixture can lead to a novel material structure: a densely packed 2-D droplet layer that acts as a glassy material, with dynamic heterogeneity that is reminiscent of cells in confluent layers \cite{angelini2011glasslike}. The packed droplet state is kinetically stabilized by local caging effects, uniquely enabled by the diversity of physically distinct phases.

Overall, our results demonstrate both a theoretical basis and an experimental realization of a highly diverse biomolecular condensate system.
Such diverse multi-phase droplet structures could be useful for the development of \emph{in vitro} spatially controlled bioreactors and fundamental physical studies of 2-D jammed states.
Further, our work lays the groundwork for exploitation of the nanostar system to investigate the behavior of multifarious systems \cite{shrinivas2021phase,jacobs2021self,chen2023programmable,teixeira2024liquid}, i.e. multi-component, multi-condensate mixtures that contain significant heterotypic interactions.

\section{Results}
\subsection{Predicting the limits of multi-phase behavior in homotypic DNA systems}
In a given set of conditions, the number of distinct, mutually immiscible phases that can be formed from a multi-component solution of self-binding NS particles is constrained by the combinatorics and energetics of the set of binding sequences that are utilized, as schematized in Fig.~\ref{fig:theory}.
Versions of these constraints apply to other condensate-forming nucleic acid systems that have been developed \cite{majumder2025sequence,Fabrini2024CoTrans,Stewart2024Modular}. 
Regarding the combinatorics, we first note that a diverse, purely homotypic mixture of NSs comprises a set of sticky ends that are both orthogonal and palindromic. 
Generally, a palindromic sequence contains an even number of bases, $M$, such that the first $M/2$ bases are complementary to the second $M/2$ bases (Fig.~\ref{fig:theory}A); given the 4 DNA bases, there are thus $4^{M/2}$ palindromes of length $M$.
We focus here on lengths $M= 4,~6,$ or $8$, which exhibit LLPS behavior in an experimentally convenient range ($20^\circ$C $<T_c <50^\circ$C) in typical salt conditions \cite{jeon2018salt,wilken2023spatial, sato2020sequence}; there are 336 unique palindromes of those lengths.

The 336 palindromes are not all orthogonal. This can be initially quantified by noting that palindromes are nested structures: even those with $M > 4$ have a 4-base palindromic core at their center (Fig.~\ref{fig:theory}A).
Accordingly, the 336 palindromes can be sorted into 16 groups according to their 4-base core sequence, with any two members of such a group being complementary at least through those core sequences.
Since 4-base complementarity can lead to relatively significant binding interactions, creating adhered and mixed phases \cite{Jeon2020sequence}, two palindromes with the same core will not be orthogonal.
This indicates that a maximum of 16 orthogonal palindromes can be created, regardless of $M$.
To prevent unwanted binding interactions, we adopt a more stringent level of orthogonality that disallows 3-base complementarity.
This reduces the maximal diversity, as limited by combinatorics, to 14, since only one core sequence of each of the pairs (ATAT, TATA) and (GCGC, CGCG) can be used (Fig.~\ref{fig:theory}C).

We further constrain the design of the set of sticky ends with the goal of having all phases share similar materials properties.
Particularly, the diverse droplets should condense at comparable temperatures (and thus share similar values of $T_c$) and have comparable dynamics, here enforced through similarity in the inverse capillary velocity $\eta/\gamma$, given viscosity $\eta$ and surface tension $\gamma$.
The inverse capillary velocity sets the timescale of material relaxation, so comparable $\eta/\gamma$ ensures all phases will relax into spherical droplets at similar rates. 

We enforce these constraints through a proxy parameter $\Delta G$, the two-state sticky end binding energy.
$\Delta G$ is relevant because, at constant NS valence and size, $T_c$ is governed by the enthalpy and entropy values, $\Delta H^\circ$ and $\Delta S^\circ$, that define $\Delta G$ \cite{rovigatti2014accurate,locatelli2018accurate}.
Further, prior work has demonstrated that NS liquid viscosity is controlled by the breaking of a single sticky end through an Arrhenius relation, $\eta \propto \exp\left(\Delta G/RT\right)$ \cite{conrad2019increasing}.
The surface tension is sensitive to $T_c$ and condensed-phase density \cite{rowlinson2013molecular,wilken2024nucleation};  since density is similar for particles of similar size, $\gamma$ is also controlled by $\Delta G$.
Thus, $\Delta G$ is relevant for all NS materials properties, and can be easily estimated using the SantaLucia DNA thermodynamics dataset \cite{santalucia2004thermodynamics}.
Further, the set of relevant palindromes spans a wide range of $\Delta G$ ($>15$~kcal/mol; Fig.~\ref{fig:theory}D), emphasizing the very different materials properties that could result from different palindrome choices. 
For example, a change in $\Delta G$ of only 1 kcal/mol is expected to change $T_c$ by more than 5$^\circ$C (Fig.~S1), and to change the viscosity 5-fold.

To enforce similar $\Delta G$ while respecting 3-base orthogonality, we carried out a graph-theoretic analysis of the set of 4-, 6-, and 8-base palindromes:
We numerically constructed a graph in which nodes represented palindromes, and connected two nodes with an edge only when the two palindromes were 3-base orthogonal and had a difference in $\Delta G$ less than a tunable spread parameter $\sigma_G$.
In such a graph, the largest fully connected subgraph (i.e. the `maximum clique') corresponds to the largest set of palindromes that, on a pairwise basis, fully satisfy both constraints; such a set is then a candidate for forming the maximally diverse set of NS liquids.

In practice, we constructed the graph using palindromes to which an extra unpaired base, T, was appended to either the 5' or 3' side, following NS design principles in which inclusion of such an unpaired base is predicted to lead to extra internal flexibility that promotes the liquid state \cite{rovigatti2014gels, smallenburg2013liquids, nguyen2017tuning}. This unpaired base also disrupts base stacking between the sticky end and the NS arm, simplifying free energy analysis.
In choosing $\sigma_G$, we considered the uncertainty in prediction of $\Delta G$ from the SantaLucia values, which, for relevant palindromes and at $25^\circ$C, we estimate to be roughly $0.4$ kcal/mol (Section S1).
It is meaningless to analyze $\Delta G$ differences on smaller scales; accordingly, we calculated the graphs for $\sigma_G$ values of $0.4, 0.8$ and $1.2$ kcal/mol, and found that the largest clique contained, respectively, 8, 9, and 10 sequences.
In each case, there were hundreds to thousands of such maximum cliques, though with partially overlapping compositions.
The set structure thus showed an expected tradeoff: higher diversity (i.e. larger maximum cliques) comes at the cost of reduced similarity in binding energy (i.e. larger $\sigma_G$).

To validate this analysis, we  selected one set of 9 sticky ends from the $\sigma_G = 0.8$ kcal/mol cliques for synthesis and experiment (see Fig.~\ref{fig:theory}E).
We analyzed the  pairwise interactions for that particular set using the NUPACK DNA assembly software package~\cite{zadeh2011nupack}, calculating the concentrations of all possible dimers in a mixture containing those 9 sticky ends at 12 $\mu$M each (matching later experiments which utilized 4-armed NSs at 3 $\mu$M) and 500 mM NaCl at $25^\circ$C. That calculation showed that the most probable unintended dimer (the 3/7 pair) had an expected concentration more than 7,000 times lower than all of the intended (palindromic) dimers; this verifies that our design strategy was sufficient to significantly favor homotypic over heterotypic interactions (Fig.~\ref{fig:theory}F).
We used the chosen set to create 9 distinct NSs, each of which was designed to self-assemble from 4 DNA oligomers into a structure with 4 double-stranded arms, joined at a junction, and terminating in one of the sticky ends.
In practice, NS self-assembly will always be imperfect, due both to errors in the synthesis of the oligomers and to inexact mixing stoichiometry of the 4 components.
Such assembly imperfections can lead to NSs with exposed single-stranded internal regions that can lead to unwanted NS-NS interactions \cite{Jeon2020sequence}.
To minimize this, we designed internal regions of each of the 9 NSs with sequences that, between NSs, had no complementarity on the level of 7 or more successive bases; full oligomer designs are given in the Supplement.

\subsection{Experimental characterization of NS phases}
\begin{figure}
  \centering
  \includegraphics[width=8.8cm]{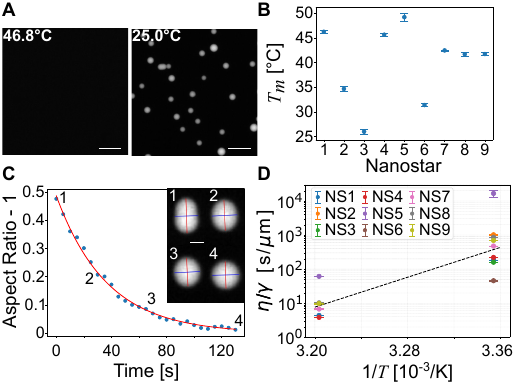}
  \caption{A) Representative images from a decreasing temperature ramp of a solution of a single type of fluorescently labeled NS (3~$\mu$M NS1, 500 mM NaCl) showing the appearance of  droplets after $T$ passes through the melting temperature, $T_m$. Scale bar: 50~$\mu$m. B) Estimates of $T_m$ for each of the 9 droplet species, obtained from the image difference variance of images such as in panel (A) (Fig.~S3). Each point represents the mean over 3 or 4 separate sample preparations; error bars: SEM. C) Representative analysis of the relaxation dynamics of a coalescing pair of droplets (shown: NS1 at $39^\circ$C), quantified through the change in aspect ratio, $A$, with time, and fit to $A-1 = C\text{e}^{-t/\tau}$. Example images are shown in the inset, with major and minor axes noted; scale bar: 10~$\mu$m. D) Estimated inverse capillary velocity of all 9 NSs, as measured from coalescence analysis (as in panel (C)), vs.~inverse temperature. Measurements were performed at $25^\circ$C and $39^\circ$C (for NSs with $T_m>$ 39$^\circ$C). 5 coalescence events were analyzed for each data point; error bars: SEM. The dashed line indicates the expected Arrhenius behavior~\cite{conrad2019increasing}, using a value for enthalpy ($\Delta H^\circ = -50$ kcal/mol) characteristic of these sequences. See Fig.~S4 for the exponential relaxation traces.
}
  \label{fig:characterizing}
\end{figure}

Each NS was assembled by mixing the 4 constituent oligomers, along with a small amount  of a fluorescently labeled version of one oligomer, in a buffer solution containing 10 mM tris-HCl (pH 7.4), 1 mM EDTA, and 250 mM NaCl. The mixture was annealed and proper NS assembly was validated via agarose gel electrophoresis (Fig.~S2).
We then used temperature-controlled fluorescence microscopy to characterize each NS's phase behavior and material properties in separate solutions (Fig.~\ref{fig:characterizing}).
Particularly, we mixed solutions of each NS at 3 $\mu$M and 500 mM NaCl, loaded each solution into a passivated glass flow cell (see Methods), and heated to $\approx55^\circ$C, which melted any existing droplets and homogenized the solution. We then subjected the solution to a slow temperature decrease at a rate of 5.1 $^\circ$C/hour while visualizing with fluorescence microscopy (Fig.~\ref{fig:characterizing}A).
We used the temperature at which droplets appeared (Fig.~S3) as an estimate of the effective melting temperature, $T_m$ (Fig.~\ref{fig:characterizing}B).
This procedure will yield a slight underestimate of the phase boundary, given delays in droplet appearance due to the timescales of nucleation and droplet growth~\cite{wilken2024nucleation}, but we expect much faster nucleation dynamics at these higher salt conditions and further expect the delays to be roughly the same for all phases, thus permitting inter-phase comparison.
We also used temperature-controlled fluorescent visualization at $T<T_m$ to analyze coalescence events, particularly quantifying their timescale of relaxation, $\tau$ (Fig.~\ref{fig:characterizing}C). This, along with the final droplet size $R$, permitted an estimate of the inverse capillary velocity for each phase as $\eta/\gamma \approx \tau/R$ (Fig.~\ref{fig:characterizing}D) \cite{leal2007advanced}.

With a few exceptions, the majority of the 9 NSs showed similar materials properties, as designed.
For example, 5 of the NSs had comparable melting temperatures around $T_{m} = 44 \pm 2^\circ$C (Fig.~\ref{fig:characterizing}B). 
Further, 7 of the NSs had $\eta/\gamma$ values that were similar, particularly clustered around $400$ s/$\mu$m at 25$^\circ$C, and decreasing to around 10 s/$\mu$m at $39^\circ$C (Fig.~\ref{fig:characterizing}D).
The variation of $\eta/\gamma$ with $T$ for most of the phases is consistent with $\eta/\gamma \propto \exp{(-\Delta H^\circ/RT)}$  (the line in Fig.~\ref{fig:characterizing}D), using $\Delta H^\circ = -50$ kcal/mol, which is typical for these sequences~\cite{santalucia2004thermodynamics}.
This Arrhenius form has been found previously to describe NS viscosity \cite{conrad2019increasing}; its success here indicates the $T$-dependence of $\eta/\gamma$ is primarily due to the viscosity, with little change of surface tension with $T$.
This is consistent with recent work \cite{wilken2024nucleation}, and likely is due to the dense phase density being relatively constant at $T\ll T_c$ in NS systems \cite{biffi2013phase,conrad2022emulsion}.
Finally, a few NSs deviated in their material behaviors: NS2, NS3, and NS6 all had significantly lower $T_{m}$, while NS5 and NS6 showed, respectively, relatively high and low $\eta/\gamma$ values (Fig.~\ref{fig:characterizing}).
The apparent mechanisms of those deviations are discussed below.

\subsection{Nine distinct coexisting NS phases}
\begin{figure}
  \begin{center}
    \includegraphics[width=\columnwidth]{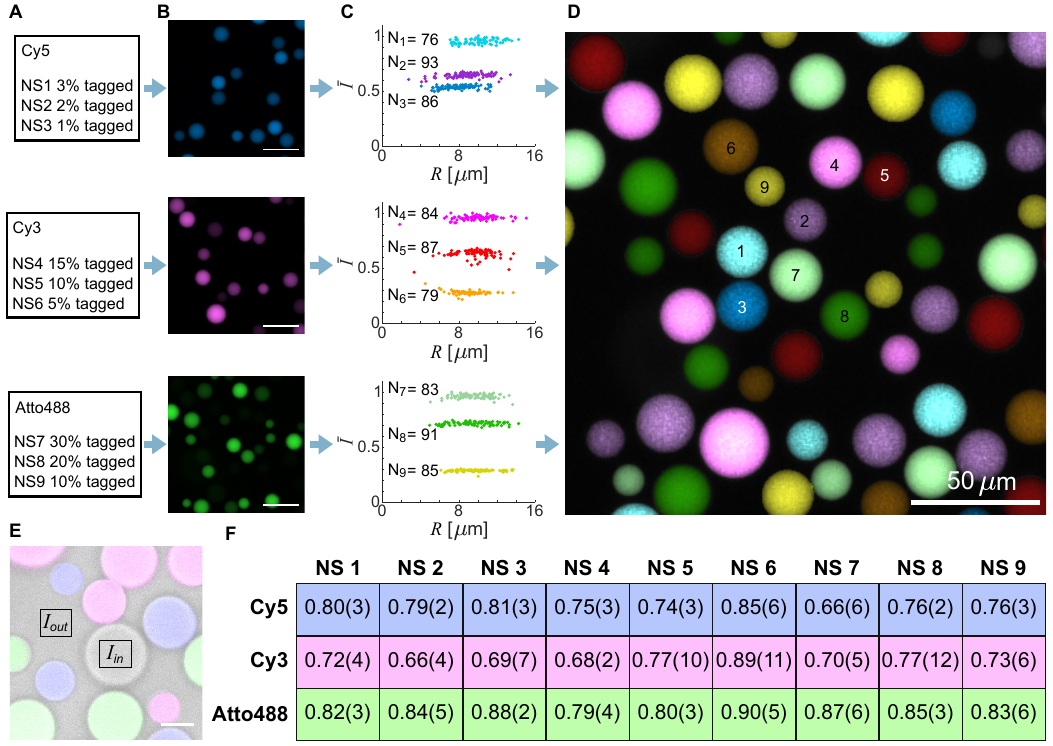}
  \end{center}
  \caption{Analysis of a solution containing all 9 NSs, at $3~\mu$M each, and 500 mM NaCl. A) 3 fluorescent dyes, at 3 concentrations each, were used to label the 9 NSs. B) The solution was imaged in 3 colors in rapid succession, showing droplets of the 3 different NSs that were visible in each channel (scale bar: 50~$\mu$m). C) Images were analyzed to find the radius, $R$, and normalized fluorescent intensity per unit volume, $\bar I$, of each droplet (Fig.~S6). Plotting these parameters revealed 3 clusters in each fluorescent channel, corresponding to the 3 NSs labeled with the relevant dye; this permitted identification of each droplet species. The number of droplets for each cluster is listed. D) The result of merging the images shown in (B), and false-coloring by droplet species, as determined by the clustering analysis; see also Fig.~S8. E) A representative cropped image from a `single dark' experiment, where NS1 is untagged, and NS2-9 are tagged per panel (A). The image is a merge of the 3 fluorescent channels, along with a bright field image to see the untagged droplet. Boxes indicate typical regions chosen to measure intensity inside the `dark' droplet, $I_{in}$, and the intensity of the background, $I_{out}$. Scale bar: 10 $\mu$m. F) Partition coefficients, $I_{in}/I_{out}$, for each color (rows), from 9 separate experiments in which the indicated NS was left unlabeled (columns). Data was averaged over 5 droplets and 5 dilute regions for each sample; the parenthetical indicates the SEM of the final digit(s).}
  \label{fig:coexisting}
\end{figure}

We proceeded to test whether the 9 NS designs could create 9 fully distinct phases in a single mixture.
We mixed solutions containing all 9 NSs, each at a concentration of 3 $\mu$M, and 500 mM NaCl.
Since most microscopes can only distinguish a few fluorescent dyes, we varied both the fluorophore identity and labeling density between NSs (Fig.~\ref{fig:coexisting}A) in order to differentiate the droplets.
Control measurements indicate $T_m$ is not sensitive to labeling (Fig.~S5), ensuring that the multiplexing scheme used was not perturbative.

The mixture of 9 NSs was loaded into a passivated flow cell, heated to $\approx55^\circ$C to homogenize the solution, and slowly cooled to 25$^\circ$C over 16.5 hours.
The solution was then allowed to sit at 25$^\circ$C for 52.5 hours; this extended incubation allowed the exceptionally viscous NS5 to round into spherical droplets.
We then captured epifluorescent images in quick succession at the 3 wavelengths of the dyes utilized (Fig.~\ref{fig:coexisting}B).
Images were analyzed by segmenting to isolate individual droplets, then fitting each droplet's radial profile to that of a sphere projected into 2-D, thus extracting an estimate of its fluorescent intensity per unit volume (Fig.~S6).
Plots of the resulting intensity density versus droplet size revealed three well-separated clusters in all three colors, corresponding to phases with, respectively, high, medium, and low labeling intensity (Fig.~\ref{fig:coexisting}C).
We used this clustering to identify each imaged droplet with a specific NS phase, permitting the generation of false-colored images with a distinct color for each phase (Fig.~\ref{fig:coexisting}D).

The image analysis shown in Fig.~\ref{fig:coexisting}A-D conclusively demonstrates that the 9 NS mixtures indeed contained 9 phases, and the lack of adhesion between droplets demonstrates that the phases were distinct from each other. 
However, this does not explicitly show that the phases are purely homotypic, i.e. that each is composed of a single type of NS.
To investigate this, we carried out `single-dark' control experiments (Fig.~\ref{fig:coexisting}E): we imaged mixtures containing all 9 NSs, with 8 of the components labeled as in Fig.~\ref{fig:coexisting}A, but with the ninth NS left unlabeled.
We mixed 9 such solutions with, respectively, each NS in turn left unlabeled, and subjected them to a slow thermal annealing procedure (see Methods). Images were then acquired using confocal microscopy in all 3 fluorescent colors and in bright field (to identify the non-fluorescent species).

The results of the 9 single-dark experiments confirmed that the mixtures were fully homotypic, and thus that the 9 phases were completely distinct with respect to both mixing and adhesion. Specifically, each experiment showed 8 fluorescent droplet species, along with one droplet species that was visible in bright field, but was, to the limits of our detection capabilities, completely dark in all 3 wavelengths.
We quantified this with a partition coefficient for each wavelength: we measured the mean fluorescent intensity within `dark' droplets, and divided by the mean intensity measured in the background spaces between droplets (Fig.~\ref{fig:coexisting}E).
The resulting 27 ratios were all smaller than 1 (Fig.~\ref{fig:coexisting}F), demonstrating that none of the 8 labeled NSs appreciably mixed into the phase formed by the unlabeled NS, for all 9 choices of unlabeled species.
In fact, the observation of ratios less than 1 indicates that the dark droplets exclude labeled NSs from the surrounding dilute solution.
The equilibrium concentration of NSs in dilute solution was, for each phase, $\approx 0.4~\mu$M (Fig.~S7), compared to the $\approx 400~\mu$M concentration of the dense phase \cite{jeon2018salt}.
Thus, the partition coefficients indicate that the dense phase contains less than 1 cross-species NS per $\approx 1000$ homotypic NSs, which is consistent with the rough estimate of cross-species interactions given in Fig.~\ref{fig:theory}F. 

Overall, the results in Fig.~\ref{fig:coexisting} demonstrate that our design indeed enables rigorous creation of a diverse and physically distinct condensate mixture.

\subsection{Diverse liquids enable dense, monodisperse droplet packings}
\begin{figure}
  \centering
  \includegraphics[width=\columnwidth]{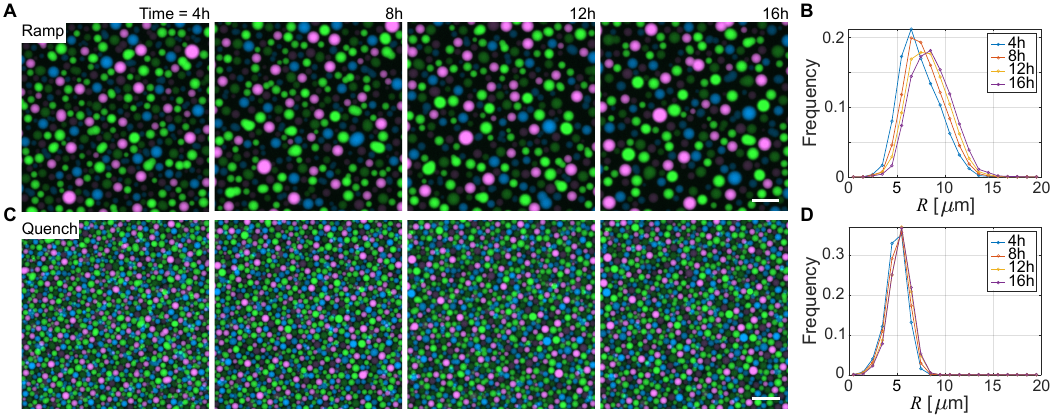}
  \caption{Thermal processing controls the droplet morphology of a 7-species NS mixture, leading to the creation of a densely packed, more monodisperse droplet layer. A and C) Portions of fluorescent images merged from 3 fluorescent channels of a 7-NS mixture after (A) a slow temperature ramp, and (C) a rapid temperature quench, from $\approx55^\circ$C to $25^\circ$C. Images shown after incubation at 25$^\circ$C for the time shown. The Cy5 channel is shown in blue, Cy3 in pink, Atto488 in green; scale bar: $50~\mu$m. Full-size images for the 16h timepoints are shown in Figs.~S10,~S11. B and D) Normalized histograms of droplet sizes were calculated from (B) post-ramp images, and (D) post-quench images for all droplets visible in the full-size images. Frequency was normalized by the number of droplets used. Mean droplet sizes are shown in Fig.~S12. Droplet sizes were calculated by fitting radial intensity profiles (Fig.~S6).
}
  \label{fig:tissue}
\end{figure}

To further quantify the diverse droplet solution, we chose to work with a reduced set of 7 NS phases, omitting NS3 and NS5, whose diverging materials behaviors led to experimental complications.
We mixed the 7 NSs at 5 $\mu$M each, in a buffer containing 500 mM NaCl and using a passivated flow cell, and subjected the solution to a similar thermal processing scheme as above, i.e. heating the mixture to the homogeneous fluid state ($\approx 55^\circ$C) and slowly cooling to $25^\circ$C at $5.1^\circ$C/hour.
We then visualized the droplets at $25^\circ$C over 16 hours and quantified their size distribution (Fig.~\ref{fig:tissue}A,B).
The droplets slowly coarsened with time and displayed a significant variation in droplet sizes (standard deviation in droplet radius: $\sigma_R = 1.96 \pm 0.02~\mu$m).
Generally, polydispersity in single-phase droplets is expected during any LLPS kinetic pathway due to the stochasticity associated with nucleation and coalescence \cite{berry2015rna}.
In this multi-phase solution, we suggest that there is another factor that contributes to polydispersity:
the differences between the phase diagrams of each species mean that, as temperature is slowly decreased, the phases with higher $T_m$ will condense earlier and have more time to coalesce and reach larger sizes, while the phases with lower $T_m$ will start forming later and have smaller sizes because coalescence is inhibited by excluded volume effects of unlike NS droplets. Supporting this, we found a correlation between droplet size and $T_m$ (Fig.~S9).
Thus the intrinsic disorder in the phase diagrams is a new source of stochasticity, beyond just nucleation and coalescence, that underlies cross-species polydispersity.

We reasoned that the effect of phase diagram disorder is amplified by the slow cooling ramp used in Fig.~\ref{fig:tissue}A, and so could be avoided by rapidly quenching the solution from the high-temperature fluid regime to a low temperature that lies within all the coexistence regimes. 
Such a quench would start droplet growth nearly simultaneously in all phases.
We thus subjected the 7-NS solution to a rapid temperature quench, from $55^\circ$C to $25^\circ$C in 35 minutes.

The rapid thermal quench resulted in a markedly different mesoscopic structure: rather than a sedimented layer of polydisperse droplets separated by `free area' of dilute solution as occurred for the slow ramp (Fig.~\ref{fig:tissue}A,B), the quench created a dense packing of smaller droplets that were more monodisperse ($\sigma_R=1.04 \pm 0.01~\mu$m), smaller droplets (Fig.~\ref{fig:tissue}C,D).
Since both processes end at the same final temperature, and near equilibrium, the total volume of the dense phase should be the same.
Yet, in this 2-D sedimented system, the area occupied by each droplet is key, and larger droplets package dense phase volume in a more area-efficient manner (e.g. one large droplet of volume $V_\circ$ takes up less area than two smaller droplets each of size $V_\circ/2$).
There is therefore a relation between droplet size and the layer morphology: the quench results in many similarly-sized droplets that are smaller, and so occupy enough area to exceed a jamming threshold, leading to the densely packed structure, while the ramp allows phases to form larger droplets that take up less area.

The mesoscopic structural difference between the ramp and quench processes also created qualitatively different dynamics. 
The droplet size distribution in the quenched system was essentially constant over a 16-hour period, exhibiting little of the coarsening seen in the ramped system (Fig.~\ref{fig:tissue}B,D).
We attribute this to the difference in the ability for droplets to coalesce, which is the dominant pathway of coarsening in NSs~\cite{wilken2024nucleation}.
The post-ramp free area permits droplets to diffuse, and so allows like-species droplets to encounter, coalesce, and coarsen. 
Conversely, the lack of free area in the densely packed system hinders such dynamics.

\begin{figure}
  \centering
  \includegraphics[width=\columnwidth]{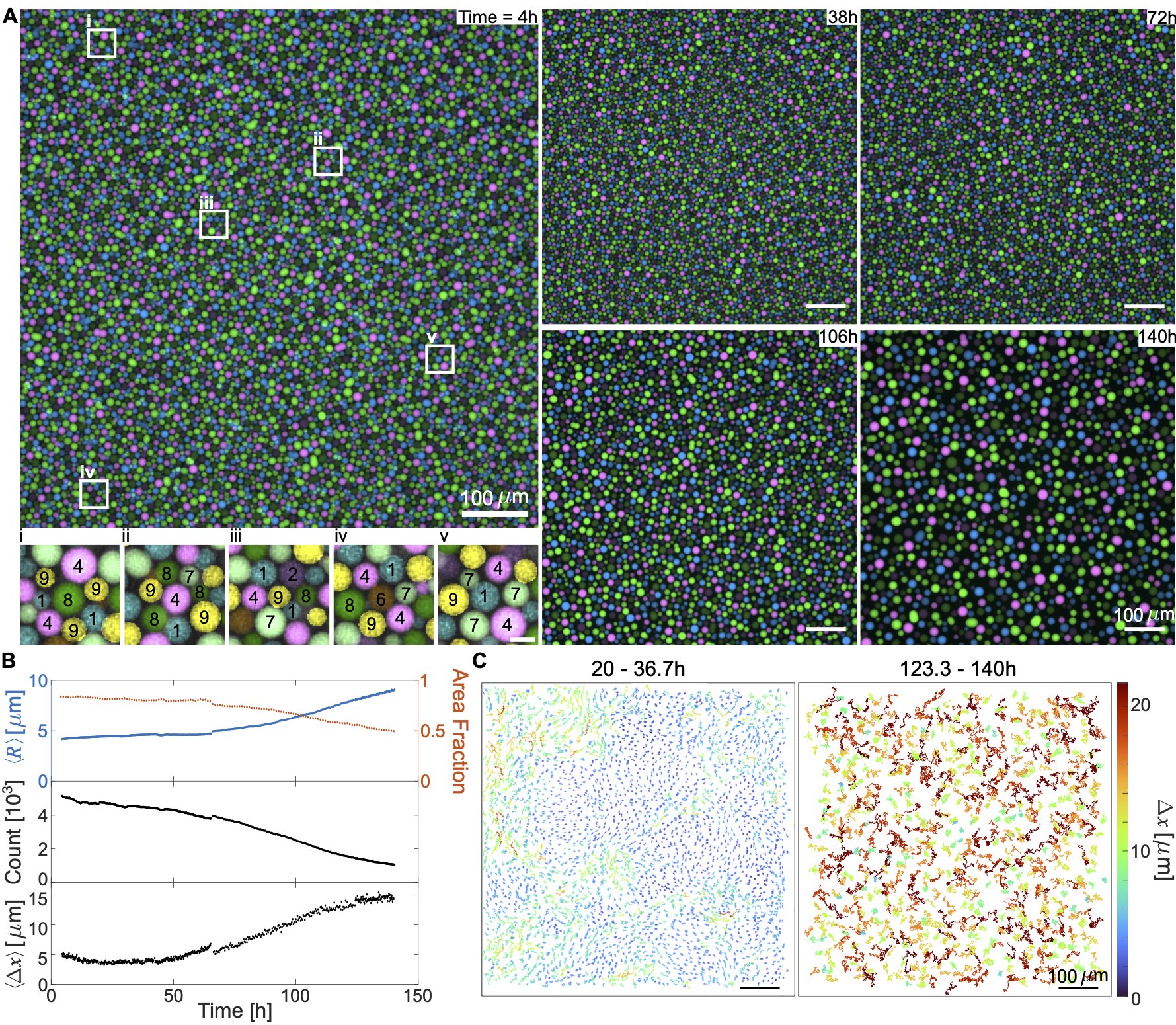}
  \caption{A) Time dependence of the morphology of 7 droplet phases after performing a temperature quench to $25^\circ$C. The Cy5 channel is shown in blue, Cy3 in pink, Atto488 in green. Insets demonstrate caging by unlike phases, and are false-colored by species (scale bar: 10 $\mu$m). B) The time dependence of mean radius, droplet area fraction, number of droplets, and mean displacement over 1 hour show initial slow dynamics, followed by faster dynamics as the system coarsens and unpacks. Discontinuities are an artifact of experimental focal drift. C) Select droplet trajectories for early (packed) and late (unpacked) stages of coarsening, colored by displacement per hour.
}
  \label{fig:trajectories}
\end{figure}

To further investigate the behavior of the quenched, densely packed system, we carried out long-term experiments in which the behavior of the 7-NS droplet layer was tracked for 140 hours, all while being held at $25^\circ$C (Fig.~\ref{fig:trajectories}). 
We analyzed the structure of the layer at early time points, finding that the dense packing creates local caged structures (Fig.~\ref{fig:trajectories}A), in which droplets of one species are tightly surrounded by those of unlike species. These cages act to trap the droplets, disallowing diffusion and coalescence, and are the mesoscopic structural mechanism underlying the nearly frozen droplet size distribution (Fig.~\ref{fig:tissue}D).

However, the long-term experiment revealed that the dense packing is not entirely static: droplets do move and coarsen, albeit in a very slow fashion, with significant increases in droplet size only occurring after $\approx70$ hours (Fig.~\ref{fig:trajectories}B).
The increasing size of droplets is correlated with a decreasing area fraction and an increasing droplet mobility over time (Fig.~\ref{fig:trajectories}B; see also Movie~S1).
This increased mobility indicates there is no droplet adhesion to the glass or to each other, confirming that the earlier slow dynamics resulted from droplet-droplet repulsions and caging.

These observations show that the quenched 7-phase system initially forms a densely packed layer, but that this layer slowly unpacks itself over time.
To investigate the mechanism underlying this behavior, we analyzed droplet trajectories (see Methods).
We found that near the end of the experiment (hours 123-140), the trajectories showed a high degree of mobility, which was relatively spatially homogeneous (Fig.~\ref{fig:trajectories}C, Movie~S1).
In contrast, near the start of the experiment (hours 20-37), the droplet trajectories were highly spatially heterogeneous, where large patches were nearly frozen while others were more dynamic (Fig.~\ref{fig:trajectories}C, Movie~S2).
The dynamic areas were associated with coalescence events that created free area, promoting cooperative motion of nearby droplets and thus further coalescence events (Movie~S3).
The quenched system, at early time points, thus displayed the dynamic heterogeneity characteristic of a jammed, glassy system of particles \cite{berthier2011dynamical}, albeit with a unique coalescence behavior.

\section{Discussion}
\subsection{Diverse and distinct liquids}
We have shown here a scheme to design and create diverse and distinct biomolecular droplets using the DNA nanostar system.
The ability to create 9 distinct, homotypic, single-component phases is likely a unique capability of a nucleic acid system, and can be attributed to the specificity of duplex hybridization.
Duplex specificity has been exploited to control a large array of interactions in the creation of rigid DNA assemblies \cite{reinhardt2016dna, ke2012three}; here we have extended this to control disordered liquid phases.
Specificity is represented by the interaction matrix of Fig.~\ref{fig:theory}F, which shows that our sticky end design achieved a rigorous level of orthogonality: the near-perfect diagonality of the matrix means that the 9 sticky ends represent a 9-dimensional vector space of interactions. 

The spectrum of orthogonal interactions exploited here likely exceeds that available to the interactions of proteins common to condensates.
Folded proteins have a wide spectrum of interaction modalities as encoded in their diverse shapes. 
However, the IDPs enriched in condensates are  thought to operate through a `sticker-spacer' paradigm, where the stickers are disordered sequences, perhaps only 1 residue in length, that drive associative interactions \cite{choi2020physical,borcherds2021how}. 
Single residues have a limited set of overlapping interaction modalities, associated with hydrophobicity, polarity, electric charge, and the ability to form disulfide bonds \cite{li1997nature}.
The limited orthogonality of protein residue interactions is likely related to observations that \emph{in vitro} reconstitutions of condensates from few-component IDP mixtures have typically only created 2 distinct phases, with significant adhesion and component-sharing between the phases \cite{Kaur2021sequence, kelley2021amphiphilic,arora2025chaperone,rai2023heterotypic,lu2025controlling,gupta2023bacterial,rana2024asymmetric,ye2024micropolarity,zhou2025multiphasic,fisher2020tunable,Feric2016coexisting}.

It is relevant to consider the diversity of condensates within the living cell: one recent work simultaneously visualized 5 condensates within the nucleus  \cite{ye2025rainbow}, and another investigated mechanisms of targeting to 12 condensates across the eukaryotic cell \cite{kilgore2025protein}, which are likely lower bounds on the actual numbers of distinct condensates present.
Such biological condensates present alternate mechanisms of diversity than those used here, e.g. by exploiting homo- and hetero-typic interactions amongst the massively diverse mixture of proteins \cite{jacobs2021self, jacobs2023theory, shrinivas2021phase, teixeira2024liquid}, or through their non-equilibrium nature, with different enzymatic processes stabilizing different phases \cite{holehouse2018functional, sawyer2019membraneless, lafontaine2021the}.
Biology also appears to utilize alternate mechanisms to maintain distinctiveness of condensates, such as amphiphilic proteins that act as stabilizing surfactants \cite{kelley2021amphiphilic}, or caging effects imposed by surrounding structures \cite{lee2021chromatin}.
Prior work has indeed incorporated some of these aspects into NS systems, such as non-equilibrium enzyme dynamics \cite{Saleh2023vacuole,Tayar2023controlling,sato2020sequence, maruyama2024temporally, Tran2023a}, or surface coatings \cite{Nguyen2019length,yamashita2024dnaorigamiarmored, gao2023controlling, Walczak2021responsive}. 
Marrying such bio-inspired features to a multi-phase NS system could extend the degree of diversity presented here and lead to new behaviors.

There are also non-biomolecular systems that form similar numbers of distinct phases as those created here.
For example, Hildebrand et al. created a mixture of atomic and small molecule liquids (including water, various oils, and liquid metals) that formed 7 phases based on differences in  polarizability and bonding modality \cite{hildebrand1949}.
Albertsson used uncontrolled stochastic polymerization to create a polymer mixture that separated into 18 distinct phases based on differences in length and monomer composition among the chains \cite{albertsson1971}.
The NS system presented here is somewhat more similar to the Hildebrand liquids, as both are based on a particulate (i.e. non-polymeric) structure that leverages orthogonal interaction modalities with only a modest number of interactions per particle.
However, the differing chemistries underlying phase stability in the Hildebrand work means the resulting liquids vary widely in materials properties, contrasting with the NS liquids, which achieve diversity while maintaining similar properties.
The Albertsson system is fundamentally different, and based on  Flory-Huggins mechanisms in which the polymers do not not have a high degree of interaction orthogonality, but separate due to differences of mixing entropy and a very large number of interactions per entangled chain \cite{rubinstein2003polymer}.
In comparison to such a polymeric system, the clear advantage of the NS system is the high degree of control, along with the biochemical specificity, which also enables partitioning of solutes to specific phases \cite{Jeon2020sequence, sato2020sequence}.
Notably, this specificity allowed us to engineer distinct (non-adhering, no intermixing of particles) liquid phases, which was not achieved in the other systems.

The DNA NS system can likely be used to create more than 9 simultaneous phases, particularly by relaxing certain constraints imposed in this work.
For example, further diversity could result by using sticky ends with more significant $\Delta G$ differences, but compensating for those differences by varying the number of arms or the size of the particles, so as to maintain materials similarity.
The principle of materials similarity itself could be discarded to maximize diversity, e.g. by creating distinct phases with behaviors that range from more solid-like to more liquid-like.
Finally, one might utilize non-palindromic sticky ends, designing either homotypic or heterotypic mixtures, which would increase the number of available sticky ends and could result in non-diagonal interaction matrices. 
Theoretical work on such heterogeneous (multifarious) mixtures indicates they might be capable of forming a diversity of multi-component (mixed) phases that even exceeds the number of components \cite{jacobs2021self}, though significant design challenges remain \cite{chen2023programmable}.

\subsection{Materials properties}
While our approach was successful in creating diverse and distinct droplets, it was only partially successful in creating identical materials properties for those droplets: of the 9 species, 7 had similar inverse capillary velocities at 25$^\circ$C, and 5 had similar $T_m$ values. The mechanisms underlying the outliers differed:
the behavior of NS6, which deviated for both $T_m$ and $\eta/\gamma$, is consistent with the duplex stability being overestimated by the nearest-neighbor sequence dataset that we utilized~\cite{santalucia2004thermodynamics}.
NS2 and NS3 both displayed a relatively low $T_m$, but similar $\eta/\gamma$, versus the others, a behavior we attribute to the ability of those sticky ends to form a self-folded (hairpin-like) state stabilized by stacking against the NS arm.
Competition with such a state is indeed expected to lower the $T_m$ \cite{santalucia2004thermodynamics}.
Further design constraints could be implemented to disallow self-folding, e.g. by altering the nature of the extra base. 

The final outlier with respect to materials properties was NS5, which showed an extraordinarily high inverse capillary velocity.
Even accounting for the exponential sensitivity of viscosity to $\Delta G$, the 40-fold higher $\eta/\gamma$ of NS5 versus the other sequences is difficult to understand.
It is unlikely to be due to the sticky end alone, since the $T_m$ was not also anomalously high.
Instead, we speculate it might be due to a unique dynamic structural behavior imparted by the internal sequence, affecting the structure or dynamics of the NS's 4-way junction or double-stranded arms.
The existence of such effects is supported by the gel mobility of NS5 (Fig.~S2) which was slower than the other particles, consistent with a more rigid internal structure.  

\subsection{Densely packed droplet layers}
Our creation of a kinetically trapped 2-D layer of droplets both illustrates interesting physical features of the diverse NS system and opens potential avenues of applied interest.
From a physical perspective, the thermal process used to create the packed layer highlights the advantages of NSs as a thermally tunable condensate system \cite{biffi2013phase}. 
The processing scheme also exploits kinetic aspects of condensate formation, with the formation of the densely packed layer enabled by the slow timescale of droplet coarsening relative to sedimentation and cage formation.
Further, the slow dynamical features of the densely packed layer are of intrinsic physical interest \cite{berthier2011dynamical}.
This behavior has parallels to the behavior of confluent epithelial cell layers, which show similar heterogeneous dynamics \cite{angelini2011glasslike}.
Notably, both systems are out of equilibrium, with morphological changes in the epithelial layers driven in part by cell division, while the droplet layer displays inverted dynamics, driven in part by coalescence events.
The multi-phase nature of the caged droplet layer raises other questions, including how each droplet species is organized within the layer, and how layer structure and dynamics might change with the number of species present.
These, and other features, are worthy of future investigation and demonstrate the rich set of physical behaviors enabled by multi-phase diversity. 

Various authors have proposed that engineered condensates in general \cite{abbas2021peptide}, and DNA condensates in particular~\cite{Udono2023dna,samanta2024dna}, have potential applications as `protocells' that act as bioreactors, drug delivery systems, or models of dynamic cellular functions.
Such applications are inspired by the ability of condensates in living cells to spatiotemporally control reactions or to sequester/exclude molecules through the differing environment inside versus outside the condensate \cite{banani2017biomolecular}.
The packed, stable 2-D droplet layer achieved here extends these concepts to highly diverse compartments spread over long distances, offering the potential to create biochemically specific condensate environments across millimeters.
The layer is a hierarchical structure, with features on the nanoscale (the specific sequence and structure of individual NSs), mesoscale (micron-scale droplets and local packings), and macroscale (the 2-D layer itself).
Thus, the 2-D droplet layer might have utility as a `prototissue' that carries out macroscopic functions that are enhanced by its substructure, which is analogous to the differentiated nature of cells within biological tissues.
Further, the layer displays long-range mechanical coupling, as evidenced by the cooperative motion of patches of droplets, indicating the possibility of using it to transmit mechanical signals, and perhaps shape changes, over long distances, in analogy to certain processes in developmental biology~\cite{Collinet2021programmed}.

\section{Conclusion}
We have presented a design scheme and experimental realization of a diverse and distinct set of biomolecular condensates using the DNA nanostar system. 
Specifically, we analyzed DNA sequence combinatorics and hybridization energetics to generate sets of palindromic sticky ends that satisfy both pairwise orthogonality and similarity in self-binding free energy.
We used one of those sets to synthesize 9 homotypic DNA NSs and tested their behavior in a common mixture, finding that they indeed created 9 distinct droplet species with neither intermixing of NSs nor adhesion between droplets. 
Further, we showed that such a multi-phase droplet mixture, after appropriate thermal processing, can form a jammed layer of droplets that displayed long-term stability due to droplet caging effects that are specifically made possible by the high diversity of the system.
We suggest that a highly diverse DNA condensate system enables the creation of novel, differentiated mesoscale structures and opens avenues of exploration of unique physical behaviors.

\section{Materials and Methods}
\subsection{Nanostar Self-Assembly}
Untagged oligos, listed in the supplement, were synthesized and desalted by Integrated DNA Technologies (IDT), resuspended in 10 mM Tris-HCl (pH 7.4) and 1 mM EDTA, filtered and reconcentrated using 100 kDa and 3 kDa centrifugal filters, and stored at 4$^\circ$C.
Tagged oligos were synthesized and HPLC purified by IDT, resuspended in 10 mM Tris-HCl, and stored frozen.
Oligos were combined in equimolar concentrations, annealed in 10 mM Tris-HCl, 1 mM EDTA, 250 mM NaCl buffer by incubating at 95$^\circ$C for 10 mins before ramping down to 4$^\circ$C at -0.5$^\circ$C/min, and stored at 4$^\circ$C.
For experiments shown in Fig.~\ref{fig:coexisting}, an additional rinse step was added using a 50 kDa centrifugal filter in order to remove NaCl from the storage buffer.

\subsection{Epifluorescent Microscopy}
Annealed NSs were mixed with other NSs (as appropriate to the experiment) in conditions of 10 mM Tris-HCl, 1 mM EDTA, and 500 mM NaCl.
The sample tube was vortexed and lightly centrifuged and heated to 55$^\circ$C to create a homogeneous mixture. The sample was pipette-mixed and loaded into an imaging chamber, which was sealed with UV curable glue. 

Glass slides and coverslips were cleaned and coated with polyacrylamide~\cite{Lau2009condensation, sanchez2013engineering} and stored at room temperature in the polyacrylamide solution. 
When ready to be used, the glass was rinsed heavily with water and gently blown dry with nitrogen gas. 
Sample channels were then constructed with parafilm as previously described~\cite{jeon2018salt, nguyen2017tuning}.

For experiments corresponding to Fig.~\ref{fig:coexisting}E, the sample was heated to $55^\circ$C and cooled to $25^\circ$C at $5.5^\circ$C/hour, with a 140 minute hold at $40^\circ$C.
The sample was held at 25$^\circ$C using a Peltier device built in house and imaged with a Crest X-Light V3 spinning disk confocal (20x objective). Images were prepared using Fiji~\cite{Schindelin2012fiji}. All other samples were placed in an Okolab UNO temperature-controlled chamber of a Nikon Eclipse Ti2-E fluorescent microscope (10x objective) and imaged using Micromanager~\cite{Edelstein2010computer}.

\subsection{Classification of Droplet Species}
Droplets were located using Otsu thresholding~\cite{crocker1996methods}, and their radii and intensity densities were extracted by fitting the radial intensity profile to the intensity expected of a sphere projected onto 2-dimensions (Fig.~S6) \cite{conrad2022emulsion, Saleh2023vacuole}.

\subsection{Coalescence Experiments}
Five coalescence events per NS phase and condition were manually identified and cropped for analysis.
Events were chosen in which droplets showed Brownian motion, and where the initial droplets were of similar size.
Cropped images containing only one ellipsoid were imported into Python, and analyzed as described in \cite{Stewart2024Modular} to track major and minor axes lengths as the droplet relaxed.
The final radius of the droplet was obtained by averaging the major and minor axes given by the last frame, except for cases where the final droplet was still an ellipsoid; for these cases, the major and minor axes were used to calculate the radius of a sphere of equivalent volume, assuming the ellipsoid was symmetric. 

\subsection{Packed Droplet Layer Analysis}
For Fig.~\ref{fig:trajectories}, droplet centroids were located by finding local intensity maxima in the images. 
The radius of each droplet was estimated by assuming a spherical profile, and comparing the intensity at the centroid to 16 nearby pixels of known separation; this method led to a slight underestimate of radius.
The area fraction was determined by training a machine learning algorithm on images throughout the 140-hour coarsening process that classified pixels as belonging to  either the dense  or dilute phase \cite{Berg2019ilastik}.

For droplet tracking, a subset of droplets were located via Otsu thresholding and tracked using custom MATLAB code based on existing methods that link droplet locations in successive frames~\cite{crocker1996methods}. To account for slow drift in the system, we subtract, from each trajectory, the average trajectory of all tracked droplets.

\section{Acknowledgements}
We thank William M. Jacobs, Michio Tateno, Anna B. N. Nguyen, Gabrielle R. Abraham, and Nathaniel Conrad for helpful discussions. This work was supported by the W.M. Keck Foundation and partially supported by the MRSEC Program of the National Science Foundation under Award No. DMR 2308708.

\bibliographystyle{apsrev4-1}
\bibliography{Biblio}
\end{document}